\documentclass{article}
\usepackage[authoryear]{natbib}
\usepackage{natbib}
\usepackage{amsmath}
\usepackage{amssymb}
\usepackage{amsthm}
\newtheorem{definition}{Definition}

\usepackage{graphicx}
\graphicspath{ {./Plots/} }
\usepackage{float}
\restylefloat*{figure}

\usepackage{xcolor}

\usepackage[colorlinks,citecolor=blue,urlcolor=blue]{hyperref}

\usepackage[linesnumbered,ruled,vlined]{algorithm2e}
\SetKwInput{KwInput}{Input}                
\SetKwInput{KwOutput}{Output}              

\usepackage{booktabs}
\usepackage{threeparttable}
\usepackage[
singlelinecheck=false 
]{caption}

\usepackage{url}

\usepackage{natbib}

\def\R{\mathbb{R}} 
\def\E{\mathbb{E}} 

\def\bt{\boldsymbol{t}}
\def\bu{\boldsymbol{u}}

\def\bx{\boldsymbol{x}}

\def\bz{\boldsymbol{z}}

\def\bC{\boldsymbol{C}}
\def\bD{\boldsymbol{D}}
\def\bE{\boldsymbol{E}}

\def\bI{\boldsymbol{I}}

\def\bP{\boldsymbol{P}}
\def\bQ{\boldsymbol{Q}}
\def\bR{\boldsymbol{R}}

\def\bU{\boldsymbol{U}}
\def\bV{\boldsymbol{V}}
\def\bW{\boldsymbol{W}}
\def\bX{\boldsymbol{X}}

\def\bZ{\boldsymbol{Z}}

\def\bSigma{\boldsymbol{\Sigma}}

\DeclareMathOperator*{\cov}{Cov}

\usepackage{setspace}
\usepackage[a4paper, total={6in, 8in}]{geometry}
\usepackage{lipsum}

\title{Truncated Gaussian copula principal component analysis with application to pediatric acute lymphoblastic leukemia patients' gut microbiome}

 \author{Lei Wang, Yang Ni\\
 Department of Statistics, Texas A\&M University\\
  Irina Gaynanova \thanks{Corresponding author:irinagn@umich.edu}\\
    Department of Biostatistics, University of Michigan
    \date{}
}

\begin{document}
  \maketitle

\author{}
\bigskip
\begin{abstract}
Increasing epidemiologic evidence suggests that the diversity and composition of the gut microbiome can predict infection risk in cancer patients. Infections remain a major cause of morbidity and mortality during chemotherapy. Analyzing microbiome data to identify associations with infection pathogenesis for proactive treatment has become a critical research focus.  However, the high-dimensional nature of the data necessitates the use of dimension-reduction methods to facilitate inference and interpretation. Traditional dimension reduction methods, which assume Gaussianity, perform poorly with skewed and zero-inflated microbiome data. To address these challenges, we propose a semiparametric principal component analysis (PCA) method based on a truncated latent Gaussian copula model that accommodates both skewness and zero inflation. Simulation studies demonstrate that the proposed method outperforms existing approaches by providing more accurate estimates of scores and loadings across various copula transformation settings. We apply our method, along with competing approaches, to gut microbiome data from pediatric patients with acute lymphoblastic leukemia. The principal scores derived from the proposed method reveal the strongest associations between pre-chemotherapy microbiome composition and adverse events during subsequent chemotherapy, offering valuable insights for improving patient outcomes.

\end{abstract}

\section{Introduction}\label{sec:Intro}

In the past two decades, there has been an increase in microbiome studies due to increasing evidence of associations between the human microbiome and overall health \citep{martin2010early,methe2012framework,vogtmann2016epidemiologic}.  A microbiome is a community of microorganisms (such as bacteria, fungi, and viruses) that inhabit a particular environment. The rapid development of sequencing technologies, especially microbial community profiling using 16S rRNA \citep{hamady2009microbial}, has made human microbiome studies more accessible due to its cost-effectiveness compared to whole-genome sequencing. Our motivating study
collected the stool microbiome data from newly diagnosed acute lymphoblastic leukemia (ALL) patients enrolled at St. Jude Children’s Research Hospital \citep{hakim2018gut}. Myelosuppression-related infections remain important causes of morbidity and mortality in children with ALL. We aim to investigate potential associations between the microbiome data collected just before chemotherapy for $n=95$ patients and the clinical adverse events of particular concern that occur during the chemotherapy courses (i.e., infection). Specifically, we want to find a lower-dimensional representation of the bacteria taxa $p=79$ (family level) that would allow physicians to differentiate patients in terms of their risk of adverse events so that they can closely monitor and proactively treat patients at the highest risk to reduce the occurrence. However, the observed non-negative sequence of counts for each family is heavily skewed and zero-inflated, presenting challenges for traditional dimension reduction approaches based on the Gaussian distributional assumption.

Principal component analysis (PCA) is one of the most widely used dimension reduction techniques. Standardized PCA uses the eigendecomposition of the Pearson sample correlation matrix $\bSigma$ of the $p$ features (with regular PCA using the sample covariance). However, measuring dependence via Pearson correlations implicitly assumes Gaussian distribution, and this assumption is violated by skewed and zero-inflated microbiome counts \citep{srivastava2010two}. Furthermore, there are two possible sources of zeros in the data:  “true zeros” (i.e., absence of taxa in samples) and  “pseudo-zeros” (i.e., the presence is below detection limit) \citep{xu2021zero}. However, PCA treats all zeros as absolute (``true zeros").

Several non-Gaussian dimension reduction methods have been proposed. 
\citet{chiquet2018variational} extend PCA using the exponential family framework and consider the Poisson-lognormal model for zero-inflated count data.  \citet{kenney2021poisson} develop Poisson measurement error corrected PCA for microbiome data, which estimates principal components of a fixed transformation of the latent Poisson means.
 \citet{zeng2022mbdenoise} propose a zero-inflated probabilistic PCA model based on zero-inflated negative binomial distribution. Besides PCA-based methods, factor analysis is a widely used alternative for dimension reduction. 
\citet{b2018glm} build a zero-inflated quasi-Poisson factor model to accommodate the zero inflation, but it assumes a fixed proportion of zeros for all samples, ignoring between-sample variation in zero inflation.
\citet{xu2021zero} developed a zero-inflated Poisson factor model for microbiome data, assuming a zero-inflated Poisson distribution with library size as an offset and a Poisson rate inversely related to inflated zero occurrences. 
\citet{mishra2022negative} propose negative binomial reduced rank factor regression to account for the overdispersed nature of the amplicon sequencing count data. All these methods are designed to take into account zero inflation. However, the underlying parametric models do not fully account for data skewness.

In the context of skewed continuous data, semiparametric extensions of PCA have been considered. \citet{ hanSemiparametricPrincipalComponent2012, 
han2013principal} propose sparse PCA for high-dimensional data following meta-elliptical distribution, leading to a scale-invariant method that is robust to outliers and violations of Gaussianity. However, the underlying methodology only provides an estimation of loading vectors and can not be used to obtain scores for dimension reduction. \citet{hanScaleInvariantSparsePCA2014} extend the methodology to enable dimension reduction, but the framework is still restricted to continuous data types, treating zeros as absolute. 

In summary, despite the significant developments, the existing methods for non-Gaussian data either rely on parametric assumptions (e.g. Poisson distribution), which do not fully capture the skewness in the data, or ignore zero inflation.
To simultaneously overcome the challenges of skewness and zero-inflation, we propose a truncated copula PCA method based on a semiparametric model for skewed zero-inflated data in \citet{yoonSparseSemiparametricCanonical2020}. The model assumes that the observed data are transformed from latent multivariate Gaussian variables, with skewness accommodated via unknown monotone transformations (copula) and zero-inflation accommodated via truncation of values below a certain threshold to zero.
While the model has been successfully used for canonical correlation analysis \citep{yoonSparseSemiparametricCanonical2020} and graphical model estimation \citep{yoon2019microbial}, it has not been adopted for dimension reduction due to the lack of principled definition of reduced dimension as well as difficulties associated with the lack of one-to-one mapping between the observed zeros and the latent data level. Here, we overcome these challenges by considering a spiked model for latent correlation matrix $\bSigma$ \citep{johnstone2001distribution}, thus providing a principled definition of reduced dimension, and by utilizing conditional expectations of truncated Gaussian distribution in estimation to connect observed zeros with the latent variables. Furthermore, we show that the proposed model is closely connected to probabilistic PCA \citep{tipping1999probabilistic}. Since the proposed estimation procedure is frequentist-based, our method is very fast compared to alternative fully Bayesian estimation approaches, such as the recent work by \citet{choi2024bayesian}.
As we finalized our method, we became aware of the concurrent work by \citet{deyFunctionalPrincipalComponent2024}, which proposes a semiparametric PCA framework based on copulas for functional data. Our work is distinct in its focus on microbiome data, rooted in connections with probabilistic PCA in matrix settings.

In our simulations, the proposed method is more accurate in the estimation of loading vectors and scores than alternative approaches.  In application to the motivating pediatric ALL patient stool microbiome data \citep{hakim2018gut}, the two leading PCA scores from our method lead to better patient clustering as measured by the average silhouette width \citep{rousseeuw1987silhouettes} than alternative approaches, and we find significant associations between the identified clusters and adverse events.

The rest of the article is organized as follows. In Section~\ref{sec:methods}, we review the truncated latent Gaussian copula model and propose the truncated copula PCA method. In Section~\ref{sec:Simulation}, we compare the performance of the proposed truncated copula PCA with other PCA and factor analysis methods in simulation studies. In Section~\ref{sec:Application}, we apply the proposed method to the motivating pediatric ALL patient stool microbiome data. In Section~\ref{sec:discussion}, we conclude with a discussion.

\section{Methodology}\label{sec:methods}

Let $\bX$ be an $n \times p$ microbiome abundance matrix, where each row is an observation and each column is a feature (e.g., phylum or genus). We assume that the observed data are zero-inflated, that is, $\bX$ contains many zeros.
Our goal is to reduce $\bX$ to a lower-dimensional matrix $\bU\in \R^{n \times r}$ with $r < p$. We call $\bU$ the matrix of scores. The scores can be used for exploratory visualization (especially with $r=2$) or in downstream analyses such as clustering and regression. 
We first review the truncated Gaussian copula model and then present the proposed model.

\subsection{Review: truncated latent Gaussian copula model}\label{subsec:Review}

Gaussian copula is a semiparametric model that allows for flexible specification of a joint distribution.

\begin{definition}\label{def:LNPN}
(Gaussian copula model). A random vector $\bx^* = (x_1^*,...,x_p^*)^T$ satisfies the Gaussian copula model, also known as a nonparanormal (NPN) model, if there exists a set of monotonically increasing transformations $f = (f_{j})^p_{j=1}$ such that $\bz = f(\bx) = \{f_{1}(x_{1}^{*}),...,f_{p}(x_p^{*})\}^{T} \sim N_{p}(\mathbf{0}, \bSigma)$ with $\bSigma_{jj} = 1$ for all j. We denote this by $\bx^* \sim NPN(\mathbf{0}, \bSigma, f)$.
\end{definition}

The truncated latent Gaussian copula model \citep{yoonSparseSemiparametricCanonical2020} is an extension of Gaussian copula models that accommodates zero inflation.  

\begin{definition}\label{def:TLNPN}
(Truncated latent Gaussian copula model). A random vector $\bx = (x_1,...,x_p)^{T} $ satisfies the truncated Gaussian copula model if there exists $\bx^* = (x_1^{*}, ...,x_p^{*})^T \sim NPN (\mathbf{0},\bSigma, f) $ such that
\begin{equation*}
x_j = I (x_j^{*} > C_j)x_j^{*}\quad(j=1,...,p),
\end{equation*}
where $I(\cdot)$ denotes the indicator function and $\bC = (C_1,...,C_p)$ is a vector of positive constants. We denote this by $\bx \sim TLNPN (\mathbf{0}, \bSigma, f, \bC)$. 
\end{definition}

The Gaussian copula model for $\bx^*$ allows for accommodating skewness of the data, whereas the additional indicator $I (x_j^{*} > C_j)$ in the truncated model allows for accommodating zero-inflation. Furthermore, the use of indicator $I (x_j^{*} > C_j)$ allows for mimicking the ``pseudo zeros" due to the sequencing technology detection limits, allowing the model to account for zeros that are not absolute.

Truncated latent Gaussian copula model has been successfully used for canonical correlation analysis with RNAseq and microRNA data \citep{yoonSparseSemiparametricCanonical2020} as well as for graphical model estimation for quantitative microbiome data \citep{yoon2019microbial}. In the case of compositional data, such as microbiome counts resulting from 16s rRNA technology, \citet{yoon2019microbial} demonstrate that the model is still appropriate for data after modified central log-ratio transformation (mclr). In short, the non-zero counts in each sample are normalized using their geometric mean, followed by log transformation and a fixed shift to preserve their relative ranking with respect to zero counts (which remain as zeros). The truncated latent Gaussian copula model is then assumed on the transformed data; we refer to \citet{yoon2019microbial} for additional details.

Despite the successful use of the truncated latent Gaussian copula framework for graphical model estimation and canonical correlation analysis, its utilization for dimension reduction is not straightforward. The main challenges are in establishing a connection between the observed data  $\bX \in \R^{n \times p}$ and latent correlation matrix $\bSigma\in \R^{p \times p}$ both in terms of providing a principled definition of ``reduced" dimension $r$ and corresponding lower-dimensional matrix of scores $\bU\in \R^{n \times r}$, and in terms of estimation when the map from observed zeros to the latent level is not one to one.

\subsection{Proposed model for dimension reduction}\label{subsec:ProposedModel}

Following Definition~\ref{def:TLNPN}, we assume that each row $\bx_i$, $i=1,\dots, n$, of $\bX$ follows independently the truncated latent Gaussian copula model with latent correlation matrix $\bSigma$. In the case of compositional data, we assume that each row has already been mclr-transformed as described in Section~\ref{subsec:Review}. Furthermore, 
we assume that the latent correlation matrix admits the following decomposition,

\begin{equation}\label{eq:Sigma}
\bSigma = \bV\mathbf{\Lambda}\bV^{T} + \sigma^2\bI,
\end{equation}
where $\mathbf{\Lambda}=\text{diag}(\lambda_1,\dots,\lambda_r)$ is a $r\times r$ diagonal matrix with positive entries sorted in descending order, $\bV\in \R^{p \times r}$ is a loading matrix with orthonormal columns, $\bI \in \R^{p\times p}$ is an identity matrix, and $\sigma^2>0$ is a residual variance term. In other words, we assume that the correlation matrix is a sum of rank-$r$ semi-positive definite matrix and a multiple of the identity matrix. Thus, the correlation matrix $\bSigma$ follows the spiked covariance matrix model with $r$ spikes \citep{johnstone2001distribution}. Specifically, from~\eqref{eq:Sigma}, the $p$ eigenvalues of $\bSigma$ are $\lambda_1 + \sigma^2 \ge \dots \ge \lambda_r + \sigma^2 > \sigma^2 = \dots = \sigma^2$, with the loadings matrix $\bV$ corresponding to the top $r$ leading eigenvectors.

Let $\bZ\in \mathbb{R}^{n\times p}$ be the corresponding latent Gaussian representation of $\bX$, that is, each row $\bx_i\in \R^p$ follows truncated latent Gaussian copula model with corresponding $\bz_i\in \R^p$ being the latent Gaussian vector with correlation matrix $\bSigma$. Given decomposition~\eqref{eq:Sigma} and rank $r$, the linear transformation that captures the most variance in $\bz_i$ is given by $\bV^{\top}\bz_i$. 
Thus, given the loading matrix $\bV\in \R^{p \times r}$, we define the matrix of principal scores $\bU\in \R^{n \times r}$ as $\bU = \bZ \bV$. In this way, our goal of finding a lower-dimensional representation of $\bX$ reduces to the problem of estimating loadings matrix $\bV\in \R^{p \times r}$ and the matrix of principal scores $\bU\in \R^{n \times r}$. In Section~\ref{subsec:PPCA}, we further show that the proposed model is closely connected to probabilistic principal component analysis \citep{tipping1999probabilistic}.

\subsection{Connection to probabilistic principal component analysis }\label{subsec:PPCA}

The proposed model \eqref{eq:Sigma} has a close connection with probabilistic PCA
(PPCA, \citealt{tipping1999probabilistic}).  
In PPCA, a $p$-dimensional random vector $\bz$ is modeled as 
\begin{equation}\label{eq:ppca}
\bz = \bW \bt +
\boldsymbol{\epsilon},
\end{equation}
where an $r$-dimensional latent vector $\bt \sim N(\mathbf{0}, \bI)$ with $r<p$, $\bW\in \R^{p \times r}$ is the rank $r$ loading matrix, and $\boldsymbol{\epsilon}\in \R^{p}$ is the isotropic Gaussian error term, that is $\boldsymbol{\epsilon} \sim N (\mathbf{0}, \sigma^2 \bI)$ with variance $\sigma^2$.
Integrating out $\bt$, the marginal distribution is $\bz \sim N(\mathbf{0}, \bSigma)$, where
\begin{equation}\label{eq:Sigma2}
\bSigma = \bW \bW^T + \sigma^2\bI.
\end{equation}
The resulting decomposition~\eqref{eq:Sigma2} matches our assumed model on the latent correlation matrix $\bSigma$ in \eqref{eq:Sigma} under the following constraints on model parameters to guarantee that $\bSigma$ has ones on the diagonal:
$$
 \Vert \mathbf{w}_{j} \Vert^2 + \sigma^2 = 1,\quad \forall 1\leq j\leq p.
 $$ 
 Thus, the loading matrix $\bV$ from the proposed model~\eqref{eq:Sigma} corresponds to the matrix of eigenvectors of $\bW\bW^{\top}$ under PPCA model~\eqref{eq:ppca}.

To further provide connection with the principal scores, consider the conditional distribution of latent $\bt$ in~\eqref{eq:ppca} given $\bz$ 
\begin{equation}\label{eq:post_dist}
\bt\vert\bz \sim N\left((\bW^{T} \bW + \sigma^2 \bI)^{-1} \bW^{T} \bz,\; \sigma^{2}(\bW^{T} \bW + \sigma^2 \bI)^{-1}\right).
\end{equation}

Using that $\bW\bW^{\top} = \bV\mathbf{\Lambda}\bV^{\top}$, with some algebra (Supplement S.1) we have 
\begin{equation}\label{eq:mean_post}
\E(\mathbf{t}|\mathbf{z})= (\bW^{T} \bW + \sigma^2 \bI)^{-1} \bW^{T} \bz
=\bR (\mathbf{\Lambda} + \sigma^{2}\mathbf{I})^{-1} \mathbf{\Lambda}^{1/2} \bV^T\bz,
\end{equation}
where $\bR\in \mathbb{O}^{r\times r}$ is an orthonormal matrix. Thus, the conditional expectation corresponds to the principal scores $\bu:=\bV^{\top}\bz$ up to scaling and rotation.

\subsection{Model estimation}

Our goal is to estimate the loading matrix $\bV\in \R^{p \times r}$ and the matrix of principal scores $\bU \in \R^{n\times r}$ using the model in Section~\ref{subsec:ProposedModel}. We first provide an overview of the overall proposed estimation procedure and then describe each step in more detail:

\begin{enumerate}
\item Use the observed $\bX$ to obtain an estimate $\widehat{\bSigma}$ of latent correlation matrix $\bSigma$.

\item Select dimension $r$ based on the eigenvalues of $\widehat{\bSigma}.$ 
Set $\widehat{\bV}$ as the top  $r$ eigenvectors of $\widehat{\bSigma}$.

\item Obtain the estimated latent data matrix $\widehat{\bZ}\in \R^{n\times p}$ based on winsorised empirical cdf of observed data for non-zero counts, and conditional mean of truncated Gaussian distribution for zero counts.
Set $\widehat{\bU} = \widehat{\bZ} \widehat{\bV}$.
\end{enumerate}

First, to estimate the latent correlation matrix $\bSigma$, we take advantage of a rank-based estimator developed by \citet{yoonSparseSemiparametricCanonical2020}. For each pair of features $j$ and $k$, the estimation approach is based on computing Kendall's $\tau$ from observed data to get $\widehat \tau_{jk}$, and then obtaining a moment-based estimator of underlying latent correlations as $\widehat{\Sigma}_{jk} = \arg\min_{r}\{F(r)-\widehat{\tau}_{jk} \}^2$, where $F(\cdot)$ is the bridge function such that the moment equation $F(\Sigma_{jk}) = \E(\widehat \tau_{jk})$ holds.
\citet{yoonSparseSemiparametricCanonical2020} derived the explicit form of the bridge function for the truncated latent Gaussian copula model.  
The obtained element-wise estimator is projected onto the cone of positve-semidefinite matrices to form $\widehat{\bSigma}_p$. The final $\widehat{\bSigma}$ is obtained from $\widehat{\bSigma}_p$ by applying shrinkage with $\upsilon>0$ as $\widehat{\bSigma} = (1- \upsilon)\widehat{\bSigma}_p + \upsilon I$ to guarantee strict positive definiteness. \citet{yoonSparseSemiparametricCanonical2020} prove that the resulting $\widehat \bSigma$ is consistent for $\bSigma$ as long as $\upsilon \leq (\log p/n)^{1/2}$.   In practice, this estimation can be done with R package \textsf{latentcor} \citep{huangLatentcorPackageEstimating2021}. We use the default value of $\nu  = 0.001$.

Second, we select dimension $r$ based on the eigenvalues of $\widehat \bSigma$.
 A simple and widely used method is to select $r$ based on the cumulative percentage of total variation explained based on a scree plot of the eigenvalues of the correlation matrix \citep{cattell1966scree}. This is the approach we adopt for the analysis of the ALL microbiome data in Section~\ref{sec:Application}. However, other approaches can also be used with our methodology. For example, the profile likelihood approach based on clustering eigenvalues into two groups \citep{zhuAutomaticDimensionalitySelection2006} or the edge distribution method \citep{onatskiDeterminingNumberFactors2010} that looks for significant eigenvalue separation based on the square root shape of the edge of the eigenvalue distribution. 
 After choosing $r$, the loading matrix $\widehat{\bV}$ is formed from the top $r$ eigenvectors of $\widehat \bSigma$.

Finally, since we define the principal score matrix as $\bU = \bZ\bV$, to obtain the estimate $\widehat{\bU}$ we need to obtain the estimate of the latent data matrix $\widehat{\bZ}$. The estimation approach depends on whether the corresponding observed $\bX$ is zero or not. For the non-zero elements of $\bX$, we estimate the underlying monotone transformation from Definition~\ref{def:LNPN} based on winsorized empirical cumulative distribution function (cdf).
For the zero elements of $\bX$, the map to the underlying latent $\bZ$ is not one-to-one; we use the conditional expectation approach by \citet{chungSparseSemiparametricDiscriminant2022}. We next provide more details for each of these steps in obtaining $\widehat{\bU}$.

Let $\bx_j\in \R^n$ be the observed $j$th truncated variable with the first $m_j$ elements being non-zero (without loss of generality). Let $F_j$ be the empirical cdf of $\bx_j$, and let
$\widehat F_j$ be the empirical cdf of $\bx_j$ winsorized at level $\delta = (2n)^{-1}$, i.e.
$$ \widehat{F_j} \left( \delta, x_{1j},...,x_{nj} \right) = W_j^{\delta} \left( \frac{1}{n} \sum_{i=1}^{n} 1(x_{ij} \le t) \right), $$ where 
$$  W_j^{\delta}(x) = 
\begin{cases}
      \delta & \text{if x} \le \delta \\
      x & \text{if } \delta < x \le 1-\delta\\
      1-\delta & \text{if x} > 1-\delta 
    \end{cases}       $$
Let $\Phi^{-1}$ be the quantile function of the standard normal distribution. Then we can obtain an estimate of the underlying monotone transformation by setting $\widehat f_j(t) = \Phi^{-1}\{\widehat F_j(t)\}$, similar idea is used in \citet{hanScaleInvariantSparsePCA2014, chungSparseSemiparametricDiscriminant2022}. Thus, we obtain $\widehat z_{ij} = \widehat f_j(x_{ij})$ for non-zero entries $x_{ij}$, $i<m_j$, by mapping them to the latent level via $\widehat f_j$.

For zero entries $x_{ij}$, $i> m_j$, from Definition~\ref{def:TLNPN}, it must hold that $x^*_{ij}\leq C_j$, and correspondingly $z_{ij} \leq \delta_j$ for latent threshold $\delta_j = f_j(C_j)$. This threshold can be estimated consistently by $\widehat \delta_j = \Phi^{-1}\{(n-m_j)/n\}$. To further obtain $\widehat z_{ij}$, we propose to use its conditional expectation based on all variables within the same sample $i$. Specifically, for sample $i$, let $\mathcal{N}_i$ indicate the variables with observed non-zero values, and $\mathcal{N}_i^c$ the observed zeros. Denote the corresponding subvectors of $\bz_i\in \R^p$ by $\bz_{i\mathcal{N}_i}$ and $\bz_{i\mathcal{N}_i^c}$. Then for zero $x_{ij}$, we propose to consider conditional expectation
\begin{align}
\E_{\bSigma}(z_{ij}| x_{i1}, \dots, x_{ip}) = \E_{\bSigma}(z_{ij}| \bz_{i\mathcal{N}_i},\ z_{ik} < \delta_k\ \mbox{for}\ k\in \mathcal{N}_i^c),
\end{align}
where subscript $\bSigma$ emphasizes that this expectation is a function of true correlation matrix $\bSigma$. Since both $\bSigma$ and latent $z_{i\mathcal{N}_i}$, $\delta_k$ are unknown, we set
\begin{align}
\widehat z_{ij} &= \E_{\widehat \bSigma}(z_{ij}| \widehat \bz_{i\mathcal{N}_i},\ z_{ik} < \widehat \delta_k\ \mbox{for}\ k\in \mathcal{N}_i^c),
\label{eq:conditional}
\end{align}
where we replace each $\delta_k$ with its consistent estimator $\widehat \delta_k$, replace $z_{il}$ corresponding to non-zero observed $x_{il}$ with $\widehat z_{il} = \widehat f_l(x_{il})$, and replace the unknown latent correlation matrix $\bSigma$ with its consistent estimator $\widehat \bSigma$ as described above.  Equation~\eqref{eq:conditional} corresponds to the expected value of truncated normal distribution. As it is not available closed-form, we evaluate it based on averaging MCMC samples from truncated normal distribution, for which off-the-shelf R packages are available, e.g. \textsf{tmvtnorm} \citep{wilhelm2022package} and \textsf{tmvnsim} \citep{bhattacjarjee2016package}. 

Finally, we obtain $\widehat \bZ \in \R^{n \times p}$ by combining all estimated $\widehat z_{ij}$, $i=1,\dots, n$, $j=1, \dots, p$, and estimate the matrix of principal scores as $\widehat{\bU} = \widehat{\bZ} \widehat{\bV}$. We refer to our method as truncated copula PCA.

\section{Simulation studies}\label{sec:Simulation}

\subsection{Methods for comparison}

We compare the proposed truncated copula PCA, with the following existing methods: (a)~standard PCA, implemented in \textsf{prcomp} function from R package \textsf{stats}; we consider both scaled and unscaled versions; (b) PoissonPCA  \citep{kenney2021poisson}, implemented in R package \textsf{PoissonPCA}; and
(c) ZIPFA \citep{xu2021zero}, zero-inflated Poisson factor model, implemented in R package \textsf{ZIPFA}. Both PossionPCA and ZIPFA are designed for count data with zero inflation, thus in our simulation, we simulate non-negative zero-inflated count data. 

\subsection{Data generation}

We consider $n=200$ independent observations and $p=20$ features. 
We first generate latent Gaussian $\bZ\in \R^{n \times p}$ following probabilistic PCA model in Section~\ref{subsec:PPCA} with $r=4$ as
$$
\bZ = \bU\bW^{T} + \bE,
$$
where all elements of score matrix $\bU\in \R^{n \times r}$ are generated independently from $N(0,1)$ and the elements of the error matrix $\bE\in \R^{n \times p}$ are generated independently from $N(0, \sigma^2)$ with $\sigma =0.1$. To generate $\bW\in \R^{p \times r}$, we first draw all elements independently from $N(0,1)$, and then normalize each row by $\sqrt{1-\sigma^2}$ to ensure that the covariance matrix of $\bSigma$ is a correlation matrix, that is, by construction,
$$
\cov(\bz_i) = \bW\bW^{\top} + \sigma^2\bI = \bSigma
$$
has 1s on the diagonal. The true loadings $\bV\in \R^{p \times r}$ are the leading $r$ eigenvectors of $\bSigma$.

Given $\bZ$, we generate $\bX^*$ (with each row following Guassian copula model from Definition~\ref{def:LNPN}) by considering four types of transformations:
\begin{enumerate}
\item Scaling transformation, $\bX^* = 10\bZ$. The transformed data still have a normal distribution but with a higher variance (to better match what is observed in real data).
\item Cubic transformation, $\bX^* = 10*{(\bZ)}^3$ applied element-wise. The transformed data no longer have a normal distribution.
\item Exponential transformation, $\bX^* = 10e^{\bZ}$ applied element-wise. The transformed data no longer have a normal distribution.
\item Mixed transformation, $\bX^* = 10 + 10\bZ + 10*{(\bZ)}^3 + 10e^{\bZ}$ applied element-wise. The transformed data no longer have a normal distribution.
\end{enumerate}

To generate zero-inflated $\bX$, we first shift $\bX^*$ by a constant so that the minimal values are zeros, i.e., we add $\min(\bX^*)$ to all elements of $\bX^*$ leading to strictly non-negative values. Subsequently, we introduce zero inflation by setting the smallest $\pi_0$ elements of $\bX^*$ to zero where $\pi_0$ is considered in the range of $\{0\%, 5\%, 10\%, 15\%, 20\%, 30\%, 40\%, 50\%, 70\%, 90\%\}$. Finally, we round the values in $\bX^*$ to the nearest integers leading to the final $\bX\in\R^{n \times p}$.

For each type of transformation and each fixed level of zero inflation $\pi_0$, we consider 100 replications.

\subsection{Performance evaluation}
For all methods, we fix $r$ at ground truth ($r=4$) and obtain estimates $\widehat \bV$ and $\widehat \bU$ of true loading matrix $\bV$ and scores $\bU$, respectively. To evaluate estimation accuracy, we use chordal distance due to its invariance to scaling and orthogonal transformations. Specifically, we calculate

 $$
 \mbox{dist}(\widehat{\bV},\bV)  = \frac{1}{\sqrt{2}} \parallel \bP_{\widehat \bV} - \bP_{\bV}\parallel_{F},
\
 \mbox{dist}(\widehat{\bU},\bU) = \frac{1}{\sqrt{2}} \parallel  \bP_{\widehat \bU} - \bP_{\bU}\parallel_{F},
 $$
where $\bP$ is the projection matrix for the corresponding subspace, and $\|\cdot\|_F$ is the Frobenius norm.
The smaller the chordal distance, the more accurate the subspace estimation. 

\subsection{Simulation results}

We discuss the results for scaling and mixed transformation. The results for cubic and exponential transformation are similar and are presented in the Supplement S.2. 

Tables \ref{t-Chordal V ide}-\ref{t-Chordal U ide} show average chordal distances (with standard errors) for loadings and scores, respectively, across 100 replications for the scaling transformation. Since the transformed data still follows Gaussian distribution, as expected, standard PCA performs well and is the best-performing method in the absence of zero inflation ($\pi_0=0$). However, as the proportion of zeros increases, the truncated copula PCA outperforms both standard PCA and other methods in loading matrix estimation. 
In score estimation, truncated copula PCA performs similarly to standard PCA, however, its performance deteriorates when the proportion of zeros is very large (e.g., over 90\%). The latter can be explained by the increased errors in the estimation of underlying transformations $f_j$ in the presence of a large number of zeros. Overall, when the transformed data are still Gaussian, truncated copula PCA does not lose much compared to standard PCA and is better than alternative methods for zero-inflated data (PoissonPCA, ZIPFA).

\begin{table*}[!t]
\begin{tabular}{@{} c c c c c c @{}}
\toprule
Zero (\%) & \shortstack{truncated \\copula PCA} &  \shortstack{Standard PCA \\ (unscaled)} &  \shortstack{Standard PCA \\ (scaled)} & PoissonPCA & ZIPFA  \\ \midrule
0\%& 0.08 ( 0.02 )& $\mathbf{0.03 ( 0.00 )}$&0.07 ( 0.02 )& $\mathbf{0.03 ( 0.00 )}$& 1.02 ( 0.01 )
\\
5\%& $\mathbf{0.08 ( 0.02 )}$& $\mathbf{0.07 ( 0.01 )}$&$\mathbf{0.08 ( 0.02 )}$& $\mathbf{0.07 ( 0.01 )}$& 1.02 ( 0.01 )
\\
10\%& $\mathbf{0.08 ( 0.02 )}$& 0.12 ( 0.02 )&$\mathbf{0.09 ( 0.01 )}$& 0.11 ( 0.02 )& 1.02 ( 0.01 )
\\
15\%& $\mathbf{0.08 ( 0.02 )}$& 0.15 ( 0.02 )&0.10 ( 0.01 )& 0.14 ( 0.02 )& 1.02 ( 0.01 )
\\
20\%& $\mathbf{0.08 ( 0.02 )}$& 0.17 ( 0.03 )&0.11 ( 0.01 )& 0.16 ( 0.03 )& 1.02 ( 0.01 )
\\
30\%& $\mathbf{0.09 ( 0.02 )}$& 0.19 ( 0.03 )&0.12 ( 0.01 )& 0.19 ( 0.03 )& 1.04 ( 0.32 )
\\
40\%& $\mathbf{0.11 ( 0.02 )}$& 0.21 ( 0.03 )&$\mathbf{0.13 ( 0.03 )}$& 0.20 ( 0.03 )& 1.32 ( 0.15 )
\\
50\%& $\mathbf{0.13 ( 0.02 )}$& 0.22 ( 0.03 )&$\mathbf{0.14 ( 0.03 )}$& 0.22 ( 0.04 )& 1.50 ( 0.11 )
\\
70\%& $\mathbf{0.19 ( 0.02 )}$& 0.32 ( 0.04 )&0.25 ( 0.03 )& 0.32 ( 0.04 )& 1.57 ( 0.11 )
\\
90\%& $\mathbf{0.45 ( 0.05 )}$& 0.89 ( 0.15 )&0.79 ( 0.13 )& 0.89 ( 0.15 )& 1.66 ( 0.08 )
\\ \hline
\end{tabular}
\caption{\label{t-Chordal V ide}Chordal distance between the true and the estimated loading matrix with scaling transformation when $n=200$, $p=20$, $r=4$. Mean (SE) are reported based on 100 replications. Bolded font indicates the best performing method (within one SE). }
\end{table*}

\begin{table*}[!t]
\begin{tabular}{@{} c c c c c c @{}}
\toprule
Zero (\%) & \shortstack{truncated\\copula PCA} &  \shortstack{Standard PCA \\ (unscaled)} &  \shortstack{Standard PCA \\ (scaled)} & PoissonPCA & ZIPFA  \\ \midrule
0\%
& $\mathbf{0.19 ( 0.04 )}$
& $\mathbf{0.17 ( 0.04 )}$&  $\mathbf{0.17 ( 0.04 )}$& 0.84 ( 0.05 )& 1.01 ( 0.01 )
\\
5\%
& $\mathbf{0.20 ( 0.04 )}$
& $\mathbf{0.23 ( 0.03 )}$&  $\mathbf{0.22 ( 0.03 )}$ & 0.83 ( 0.05 )& 1.03 ( 0.01 )
\\
10\%
& $\mathbf{0.21 ( 0.03 )}$
& 0.29 ( 0.03 ) &  0.28 ( 0.03 ) & 0.83 ( 0.06 )& 1.04 ( 0.01 )
\\
15\%
& $\mathbf{0.22 ( 0.03 )}$
& 0.35 ( 0.03 ) & 0.33 ( 0.02 ) & 0.82 ( 0.06 )& 1.04 ( 0.01 )
\\
20\%
& $\mathbf{0.22 ( 0.03 )}$
& 0.39 ( 0.03 ) & 0.38 ( 0.02 ) & 0.81 ( 0.06 )& 1.05 ( 0.01 )
\\
30\%
& $\mathbf{0.24 ( 0.03 )}$
& 0.49 ( 0.02 ) & 0.48 ( 0.02 ) & 0.80 ( 0.06 )& 1.08 ( 0.03 )
\\
40\%
& $\mathbf{0.27 ( 0.03 )}$
& 0.59 ( 0.02 ) & 0.57 ( 0.02 ) & 0.81 ( 0.05 )& 1.45 ( 0.18 )
\\
50\%
& $\mathbf{0.29 ( 0.03 )}$
& 0.65 ( 0.02 ) & 0.63 ( 0.02 ) & 0.81 ( 0.05 )& 1.75 ( 0.12 )\\
70\%
& $\mathbf{0.40 ( 0.03 )}$
& 0.69 ( 0.03 ) & 0.67 ( 0.02 ) & 0.82 ( 0.05 )& 1.97 ( 0.06 )
\\
90\% & $\mathbf{1.00 ( 0.03 )}$
& 1.29 ( 0.07 ) & 1.24 ( 0.06 ) & 1.32 ( 0.05 )& 1.90 ( 0.06 )
\\ \hline
\end{tabular}
\caption{\label{t-Chordal U ide}Chordal distance between the true and the estimated score matrix with scaling transformation when $n=200$, $p=20$, $r=4$. Mean (SE) are reported based on 100 replications. Results within 1SE of the minimum distance are bolded.}
\end{table*}

Tables~\ref{t-Chordal V mixed}-\ref{t-Chordal U mixed} show average chordal distances (with standard errors) for loadings and scores, respectively, across 100 replications for the mixed transformation. Compared to scaling transformation, all methods perform worse except for the proposed truncated copula PCA. This is expected since our method is invariant to monotone transformations of the data. 
The proposed method has the smallest errors in both loading and score estiamtion across all settings. 
All methods have high error rates in estimating the scores in the presence of a large number of zeros. The results for cubic and exponential transformation are similar to those for mixed; see the Supplement S.2. for details.

Overall, the proposed truncated copula PCA is the best-performing method in the estimation of both scores and loading matrices regardless of zero inflation proportion and transformation type. In the presence of a large amount of zero inflation (over 90\%), accurate estimation of scores becomes challenging, with all methods demonstrating high error rates.

\begin{table*}[!t]
\begin{tabular}{@{}cccccc@{}}
\toprule
Zero (\%) & \shortstack{truncated\\copula PCA} &  \shortstack{Standard PCA \\(unscaled)} &  \shortstack{Standard PCA \\ (scaled)} & PoissonPCA & ZIPFA  \\ \midrule
0\%
& $\mathbf{0.08 ( 0.02 )}$& 0.29 ( 0.07 )&0.17 ( 0.04 )& 0.29 ( 0.07 )& 1.09 ( 0.03 )
\\
5\%
& $\mathbf{0.08 ( 0.02 )}$& 0.38 ( 0.10 )&0.18 ( 0.03 )& 0.38 ( 0.10 )& 1.11 ( 0.03 )
\\
10\%
& $\mathbf{0.08 ( 0.02 )}$& 0.46 ( 0.12 )&0.18 ( 0.02 )& 0.46 ( 0.12 )& 1.11 ( 0.04 )
\\
15\%
& $\mathbf{0.08 ( 0.02 )}$& 0.51 ( 0.13 )&0.17 ( 0.02 )& 0.50 ( 0.13 )& 1.12 ( 0.04 )
\\
20\%
& $\mathbf{0.08 ( 0.02 )}$& 0.53 ( 0.13 )&0.17 ( 0.02 )& 0.53 ( 0.13 )& 1.16 ( 0.07 )
\\
30\%
& $\mathbf{0.09 ( 0.02 )}$& 0.54 ( 0.12 )&0.15 ( 0.02 )& 0.53 ( 0.12 )& 1.40 ( 0.14 )
\\
40\%
& $\mathbf{0.11 ( 0.02 )}$& 0.53 ( 0.11 )&0.15 ( 0.02 )& 0.53 ( 0.11 )& 1.53 ( 0.12 )
\\
50\%
& $\mathbf{0.13 ( 0.02 )}$& 0.54 ( 0.11 )&$\mathbf{0.15 ( 0.02 )}$& 0.54 ( 0.11 )& 1.52 ( 0.10 )
\\
70\%
& $\mathbf{0.19 ( 0.02 )}$& 0.67 ( 0.15 )&0.28 ( 0.03 )& 0.67 ( 0.15 )& 1.48 ( 0.19 )
\\
90\%& $\mathbf{0.45 ( 0.05 )}$& 1.14 ( 0.12 )&0.86 ( 0.13 )& 1.14 ( 0.12 )& 1.66 ( 0.14 )
\\ \bottomrule
\end{tabular}
\caption{\label{t-Chordal V mixed}Chordal distance between the true and the estimated loading matrix with mixed transformation when $n=200$, $p=20$, $r=4$. Mean (SE) are reported based on 100 replications. Results within 1SE of the minimum distance are bolded.}
\end{table*}

\begin{table*}[!t]
\begin{tabular}{@{} c c c c c c@{}}
\toprule
Zero (\%) & \shortstack{truncated \\copula PCA} & \shortstack{Standard PCA\\(unscaled)}&  \shortstack{Standard PCA\\(scaled)} & PoissonPCA & ZIPFA  \\ \midrule
0\%
& $\mathbf{0.19 ( 0.03 )}$& 0.79 ( 0.08 )&0.74 ( 0.06 )& 1.16 ( 0.05 )& 1.12 ( 0.02 )
\\
5\%
& $\mathbf{0.19 ( 0.03 )}$
& 0.84 ( 0.07 )&0.78 ( 0.05 )& 1.18 ( 0.06 )& 1.12 ( 0.02 )
\\
10\%
& $\mathbf{0.20 ( 0.03 )}$
& 0.78 ( 0.07 )&0.69 ( 0.03 )& 1.13 ( 0.07 )& 1.15 ( 0.03 )
\\
15\%
& $\mathbf{0.21 ( 0.03 )}$
& 0.70 ( 0.08 )& 0.60 ( 0.03 )& 1.07 ( 0.08 )& 1.18 ( 0.04 )
\\
20\%
& $\mathbf{0.22 ( 0.03 )}$
& 0.65 ( 0.08 )&0.53 ( 0.03 )& 1.02 ( 0.09 )& 1.22 ( 0.07 )
\\
30\%
& $\mathbf{0.24 ( 0.03 )}$
& 0.61 ( 0.07 )&0.49 ( 0.02 )& 0.95 ( 0.10 )& 1.48 ( 0.17 )
\\
40\%
& $\mathbf{0.26 ( 0.03 )}$
& 0.64 ( 0.06 )& 0.54 ( 0.03 ) & 0.92 ( 0.10 )& 1.70 ( 0.13 )
\\
50\%
& $\mathbf{0.29 ( 0.03 )}$
& 0.68 ( 0.05 )& 0.59 ( 0.03 ) & 0.91 ( 0.09 )& 1.76 ( 0.09 )
\\
70\%
& $\mathbf{0.40 ( 0.03 )}$
& 0.79 ( 0.10 )& 0.63 ( 0.02 ) & 0.95 ( 0.11 )& 1.96 ( 0.07 )
\\
90\%& $\mathbf{1.00 ( 0.04 )}$
& 1.41 ( 0.06 )& 1.29 ( 0.06 ) & 1.43 ( 0.05 )& 1.88 ( 0.06 )
\\ \bottomrule
\end{tabular}
\caption{\label{t-Chordal U mixed}Chordal distance between the true and the estimated score matrix with mixed transformation when $n=200$, $p=20$, $r=4$. Mean (SE) are reported based on 100 replications. Results within 1SE of the minimum distance are bolded.}
\end{table*}

\section{Application to Microbiome Data}\label{sec:Application}
We apply the proposed truncated copula PCA model together with standard PCA, Poisson PCA, and ZIPFA to stool microbiome data from newly diagnosed acute lymphoblastic leukemia (ALL) patients enrolled at St. Jude Children's Research Hospital between January 2012 and August 2015 \citep{hakim2018gut}. Patients underwent chemotherapy, which consisted of a 6-week remission induction, a 8-week consolidation, and a 120-week continuation phase
that included two 3-week periods of intensive chemotherapy (reinduction I in weeks 7-9 and reinduction II in weeks 17-20). Our goal is to explore the association between microbiome data before chemotherapy and clinical adverse events associated with chemotherapy, including fever, bacteremia, neutropenia, and diarrhea. This exploration could potentially lead to proactive treatment selection based on the stool microbiome before chemotherapy begins.  

We consider $n=95$ stool samples that were collected just before the chemotherapy started and group the 6018 measured OTUs into $p=79$ family level data. We aim to apply PCA to reduce the dimensionality and use the resulting principal components to model the relationship between adverse events occurring in chemotherapy and microbiome composition before chemotherapy starts. Specifically, logistic regression is used considering the binary characteristic of adverse events.

First, the raw count data at the family level are converted into compositional data by dividing each family OTU by the total count of each sample. Modified central log ratio (mclr) transformation \citep{yoon2019microbial} is then applied to the compositional data for truncated copula PCA. Standard log-ratio transformations are applied to the compositional data for standard PCA and PoissonPCA. For ZIPFA, we use the raw count data directly. In light of the simulation results from Section~\ref{sec:Simulation}, which show all methods performing poorly with extreme zero inflation, we exclude families with 90\% or higher zero inflation, leaving  40 families for subsequent analysis. The zero inflation in these families ranges from 0\% to 86\%, with a median of $36.32\%$ and interquartile range of $(10.00\%, 65.00\%)$. 

We apply each PCA method—truncated copula PCA (proposed), standard PCA (scaled and unscaled), Poisson PCA, and ZIPFA—and use the resulting principal component scores in downstream analyses. Specifically, we fit logistic regression models using the principal component scores to estimate the risk of four adverse events during induction therapy: bacteremia, diarrhea, fever,  and neutropenia. All models adjust for patient's age, race, gender, and a binary pre-chemotherapy risk group determined by clinicians based on genotype mutations and clinical laboratory results; patients in risk group 1 are expected to have worse outcomes than those in group 0 (low risk). For each PCA method, we include the first two principal components as predictors and assess statistical significance using a p-value threshold of 0.1, reflecting the limited sample size. To compare performance across methods, we also fit models using five principal components and evaluate model fit based on Akaike Information Criterion (AIC), with lower AIC values indicating better model performance.

Neither model showed significant associations with bacteremia or diarrhea. Based on AIC comparisons, the model using two principal component scores from the proposed truncated copula PCA outperformed models based on other PCA methods, as well as models using an extended set of five principal components, for predicting neutropenia and fever (Supplement S.3).

Table \ref{t-pvalues for logistic models for Fever w 2PCs} presents predictor p-values from each model for fever with two PCs. Only the second principal component (PC2) from the proposed truncated copula PCA is significantly associated with fever ($p = 0.024$), with a positive relationship (Supplement S.3). In contrast, none of the principal components from the other PCA methods show significant associations with fever.

\begin{table}[!t]
\centering
\begin{tabular}{@{} c c c c c c @{}}
\toprule
 & \shortstack{truncated\\ copula PCA}  & \shortstack{standard PCA\\(unscaled)} & \shortstack{standard PCA\\(scaled)} & Poisson PCA & ZIPFA \\ \midrule
PC1 & 0.649 & 0.397 & 0.479 & 0.380 & 0.126 \\
PC2 & $\mathbf{0.024}$ & 0.520 & 0.137 & 0.537 & 0.216 \\
race & 0.291 & 0.419 & 0.348 & 0.419 & 0.414 \\
age & $\mathbf{0.054}$ & 0.097 & 0.079 & 0.099 & 0.130 \\
gender & 0.269 & 0.222 & 0.189 & 0.223 & 0.242 \\
risk & 0.779 & 0.731 & 0.693 & 0.737 & 0.728 \\
\bottomrule
\end{tabular}
\caption{\label{t-pvalues for logistic models for Fever w 2PCs} Logistic regression model p-values for association between fever and the first two principal components from each PCA method, adjusting for age, race, gender, and risk group. Associations considered statistically significant at the 0.1 level are bolded.}
\end{table}

Table \ref{t-pvalues for logistic models for Neutropenia w 2PCs} reports predictor p-values from each model for neutropenia with two PCs. The second principal component (PC2) from the proposed truncated copula PCA and the first principal component (PC1) from ZIPFA are significantly associated with neutropenia ($p < 0.1$), with p-values of 0.065 and 0.072, respectively. Supplement S.3 provides full model details for truncated copula PCA and ZIPFA. In both cases, the significant principal component is positively associated with the occurrence of neutropenia.

\begin{table}[!t]
\centering
\begin{tabular}{@{} c c c c c c @{}}
\toprule
 & \shortstack{truncated\\ copula PCA}  & \shortstack{standard PCA\\(unscaled)} & \shortstack{standard PCA\\(scaled)} & Poisson PCA & ZIPFA \\ \midrule
PC1 & 0.814 & 0.472 & 0.569 & 0.460 & $\mathbf{0.072}$ \\
PC2 & $\mathbf{0.065}$ & 0.775 & 0.330 & 0.830 & 0.142 \\
race & 0.267 & 0.395 & 0.340 & 0.404 & 0.314 \\
age & 0.110 & 0.165 & 0.143 & 0.168 & 0.228 \\
gender & 0.175 & 0.159 & 0.140 & 0.162 & 0.153 \\
risk & 0.577 & 0.565 & 0.533 & 0.570 & 0.547 \\
\bottomrule
\end{tabular}
\caption{\label{t-pvalues for logistic models for Neutropenia w 2PCs} Logistic regression model p-values for association between neutropenia and the first two principal components from each PCA method, adjusting for age, race, gender, and risk group. Associations considered statistically significant at the 0.1 level are bolded.}
\end{table}

\begin{figure}[!t]
    \begin{minipage}[b]{.475\textwidth}
    \includegraphics[width=1\textwidth]{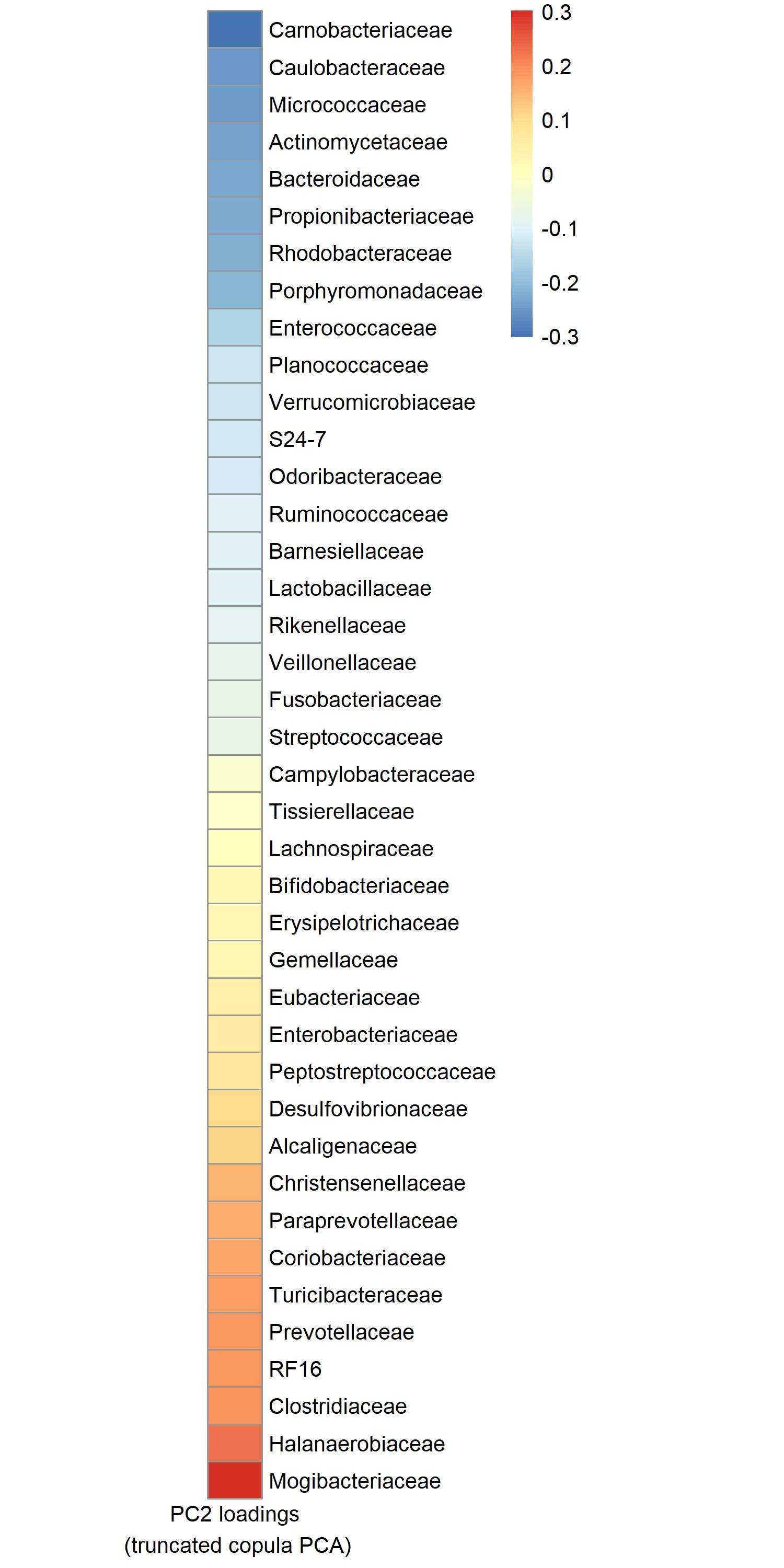}
	\caption{Heatmap for loading vector corresponding to the second principal scores from truncated copula PCA.}
    \label{fig:heatmap_V2tcPCA}
    \end{minipage}
    \hfill
    \begin{minipage}[b]{.475\textwidth}
    \includegraphics[width=1\textwidth]{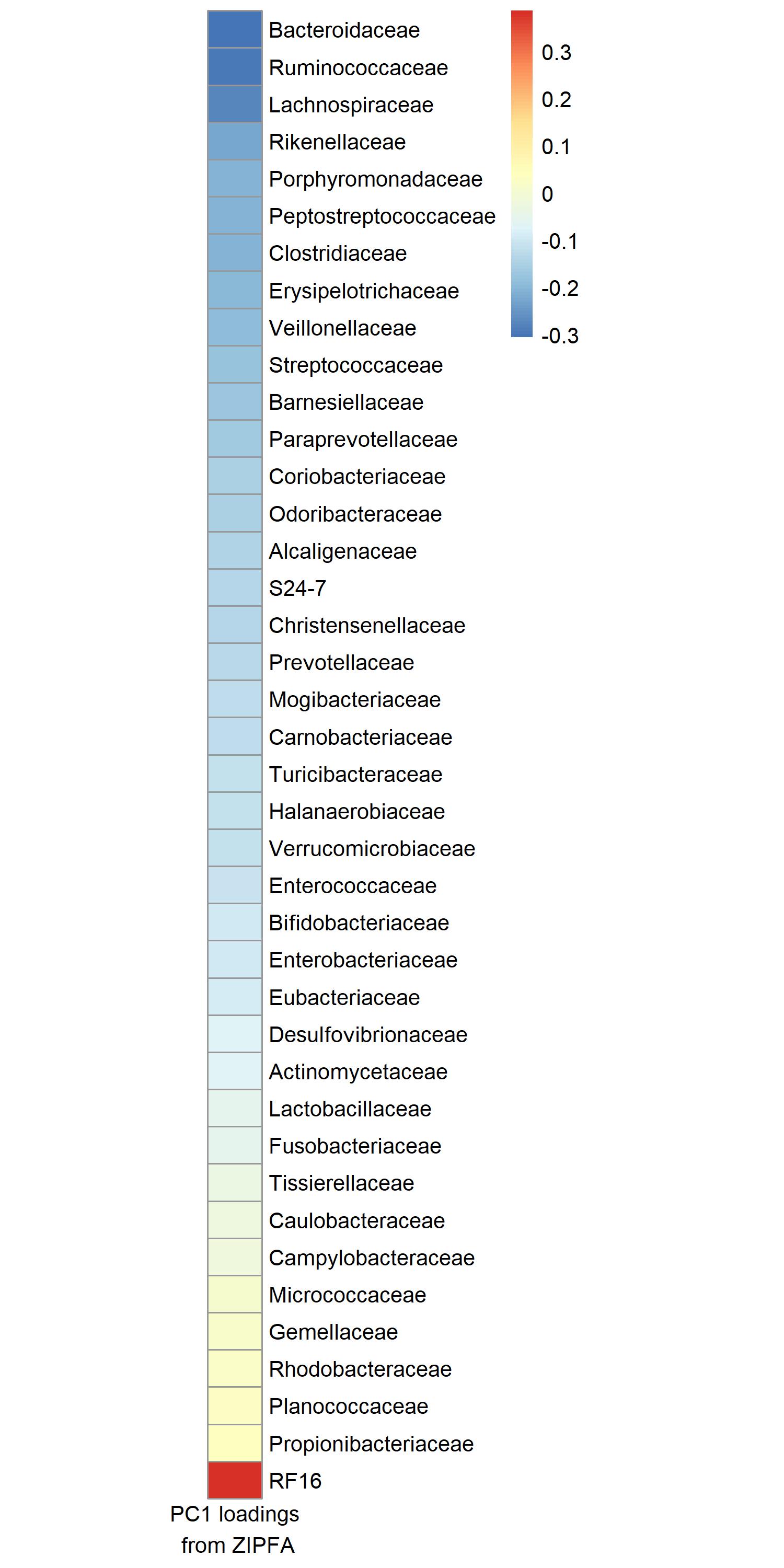}
    \caption{Heatmap for loading vector corresponding to the first principal scores from ZIPFA.}
    \label{fig:heatmap_V1ZIPFA}
    \end{minipage}
\end{figure}

Based on the results from logistic regression, only principal components identified by the proposed truncated PCA and ZIPFA showed significance for at least one of the adverse events. Specifically, ZIPFA PC1 showed positive association with neutropenia, whereas truncated copula PC2 showed positive association with both neutropenia and fever, being more comprehensive. 

Figures~\ref{fig:heatmap_V2tcPCA}-\ref{fig:heatmap_V1ZIPFA} visualize the heatmap of loadings corresponding to these components. The patterns are quite distinct, with the exception of RF16, which has large positive loading in both components,
meaning that higher levels of RF16 lead to increased risk of neutropenia (recall that the regression coefficients are positive). As the existing literature on RF16 appears to be scarce, it may be a promising target for future investigations.  Other families with high loadings in truncated copula PC2 have more established connections: \emph{Mogibacterium} has been reported to be enriched in patients with febrile neutropenia \citep{murthy2020baseline}; \emph{Clostridium difficile} and \emph{Clostridium tertium} have been linked to severe complications such as neutropenic enterocolitis, particularly in immunocompromised individuals \citep{gorschluter2001clostridium}; and members of the \emph{Prevotellaceae} family have been associated with gut dysbiosis, inflammation, and fever in the context of infections \citep{iljazovic2021perturbation}. These associations suggest that the truncated copula PCA may reveal clinically meaningful microbial patterns that are not captured by ZIPFA or other PCA methods.

\section{Discussion}\label{sec:discussion}
In this work, we developed a new truncated copula PCA method for dimension reduction of skewed and zero-inflated data. The simulation results confirm that our method is more accurate than alternative approaches in the estimation of both loading matrices and scores in the presence of a wide range of zero inflation. In the presence of high zero inflation, e.g., over $90\%$, 
the estimation of the scores matrix becomes very challenging for all methods, suggesting caution in using estimated scores in practice when zero inflation is too high. 
In our application, the proposed approach revealed that principal components derived from gut microbiome samples before initiating chemotherapy may help identify patients at risk of adverse events earlier, leading to proactive treatment selection and closer monitoring to reduce associated morbidity and mortality. 
While we have focused on applying the method to zero-inflated microbiome data, the proposed method is general and can be applied to many other data types with zero-inflation and skewness, such as microRNA data and single-cell sequencing data.

\section*{Acknowledgments}
We gratefully acknowledge Dr. Jason W. Rosch, Dr. Stanley Pounds, Dr. Li Tang, and Dr. Yilun Sun from St. Jude Children's research hospital for providing access to the microbiome data used in the application section of this study.

\section*{Declaration of conflicting interests}
The author(s) declared no potential conflicts of interest with respect to the research, authorship and/or publication of this article.

\section*{Funding}
IG was supported by NSF DMS-2422478. YN was supported by NIH 1R01GM148974-01, NSF DMS-2112943, and CPRIT RP230204.


\newpage
\renewcommand{\thesubsection}{S\arabic{subsection}}
\renewcommand{\theequation}{S\arabic{equation}}
\renewcommand{\thefigure}{S\arabic{figure}}
\renewcommand{\thetable}{S\arabic{table}}
\setcounter{equation}{0}
\setcounter{figure}{0}
\setcounter{table}{0}
\setcounter{page}{1}

\begin{center}
{\large\bf Supplementary Materials for ``Truncated Gaussian copula principal component analysis with application to pediatric acute lymphoblastic leukemia  patients' gut microbiome"} 
\medskip
\end{center}

Section~\ref{app:formulaproofs} is the proof of equation (5) in Section 2.3 in the main paper. Section~\ref{app:extrasim} includes additional simulation results for cubic transformation and exponential transformation.  Section~\ref{app:extraSJ} includes additional results for application on micro-biome data.

\subsection{Proofs equation (5) from Section 2.3}\label{app:formulaproofs}
To further provide connection for principal scores from the proposed model to the PPCA model, let's consider the conditional distribution of latent $\bt$ given $\bz$ 

\begin{equation*}
\bt\vert\bz \sim N\left((\bW^{T} \bW + \sigma^2 \bI)^{-1} \bW^{T} \bz,\; \sigma^{2}(\bW^{T} \bW + \sigma^2 \bI)^{-1}\right).
\end{equation*}

Using singular value decomposition $\bW$ can be written as $\bW = \bQ \bD \bR^T$. Linking the assumed latent correlation matrix in the proposed truncated copula PCA model, which admits 
\begin{equation*}
\bSigma = \bV\mathbf{\Lambda}\bV^{T} + \sigma^2\bI,
\end{equation*}
to the constrained PPCA model correlation matrix 
\begin{equation*}
\bSigma = \bW \bW^T + \sigma^2\bI
\end{equation*}
we get
$$
\bW \bW^{T} = \bQ \bD \bR^{T}\bR \bD\bQ^{T} = \bQ \bD^{2} \bQ^{T} =\bV \mathbf{\Lambda} \bV^{T},
$$
which implies $\bQ=\bV$ and $\mathbf{\Lambda} = \bD^2$. Further, $$\bW^{T} \bW = \bR\bD^{2}\bR^{T}.$$ 
With some algebra we have 
\begin{equation*}
\E(\mathbf{t}|\mathbf{z}) 
=(\bW^{T} \bW + \sigma^2 \bI)^{-1} \bW^{T} \bz 
=\bR (\mathbf{\Lambda} + \sigma^{2}\mathbf{I})^{-1} \mathbf{\Lambda}^{1/2} \bV^T\bz,    
\end{equation*}
where $\bR\in \mathbb{O}^{r\times r}$ is an orthonormal matrix. Thus, the conditional expectation corresponds to the principal scores $\bu:=\bV^{\top}\bz$ up to scaling and rotation.

\subsection{Additional simulation results}\label{app:extrasim}

\begin{table*}[!t]
\begin{tabular}{@{} c c c c c c @{}}
\toprule
Zero (\%) & \shortstack{truncated\\copula PCA} &  \shortstack{Standard PCA\\ (unscaled)} &  \shortstack{Standard PCA\\(scaled)} & PoissonPCA & ZIPFA  \\ \midrule
0\%
& $\mathbf{0.08 ( 0.02 )}$& 0.38 ( 0.07 )&0.24 ( 0.04 )& 0.36 ( 0.07 )& 1.12 ( 0.03 )
\\
5\%
& $\mathbf{0.08 ( 0.01 )}$& 0.67 ( 0.17 )&0.26 ( 0.03 )& 0.65 ( 0.16 )& 1.16 ( 0.04 )
\\
10\%
& $\mathbf{0.08 ( 0.01 )}$& 0.76 ( 0.17 )&0.24 ( 0.03 )& 0.75 ( 0.17 )& 1.20 ( 0.06 )
\\
15\%
& $\mathbf{0.09 ( 0.01 )}$& 0.77 ( 0.16 )&0.23 ( 0.03 )& 0.77 ( 0.16 )& 1.26 ( 0.11 )
\\
20\%
& $\mathbf{0.09 ( 0.01 )}$& 0.76 ( 0.15 )&0.22 ( 0.02 )& 0.76 ( 0.15 )& 1.40 ( 0.13 )
\\
30\%
& $\mathbf{0.10 ( 0.01 )}$& 0.72 ( 0.14 )&0.18 ( 0.02 )& 0.71 ( 0.14 )& 1.53 ( 0.11 )\\
40\%
& $\mathbf{0.11 ( 0.02 )}$& 0.68 ( 0.13 )&0.16 ( 0.02 )& 0.68 ( 0.13 )& 1.55 ( 0.10 )
\\
50\%
& $\mathbf{0.13 ( 0.02 )}$& 0.68 ( 0.13 )&0.16 ( 0.02 )& 0.68 ( 0.13 )& 1.54 ( 0.08 )
\\
70\%
& $\mathbf{0.19 ( 0.02 )}$& 0.81 ( 0.15 )&0.27 ( 0.03 )& 0.80 ( 0.15 )& 1.58 ( 0.18 )
\\
90\%& $\mathbf{0.45 ( 0.05 )}$& 1.21 ( 0.11 )&0.84 ( 0.14 )& 1.21 ( 0.11)& 1.67 ( 0.14 )\\ \bottomrule
\end{tabular}
\caption{\label{t-Chordal V cub}Chordal distance between the true and the estimated loading matrix with cubic transformation when $n=200$, $p=20$, $r=4$. Mean (SE) are reported based on 100 replications.}
\end{table*}

\begin{table*}[!t]
\begin{tabular}{@{} c c c c c c @{}}
\toprule
Zero (\%) & \shortstack{truncated\\copula PCA} &  \shortstack{Standard PCA \\ (unscaled)} &  \shortstack{Standard PCA \\ (scaled)} & PoissonPCA & ZIPFA  \\ \midrule
0\%
& $\mathbf{0.25 ( 0.03 )}$& 1.06 ( 0.06 )&1.02 ( 0.06 )& 1.33 ( 0.04 )& 1.27 ( 0.03 )
\\
5\%
& $\mathbf{0.26 ( 0.03 )}$& 1.14 ( 0.09 )&1.01 ( 0.03 )& 1.37 ( 0.06 )& 1.27 ( 0.03 )
\\
10\%
& $\mathbf{0.27 ( 0.03 )}$& 1.05 ( 0.09 )&0.87 ( 0.03 )& 1.29 ( 0.06 )& 1.33 ( 0.05 )
\\
15\%
& $\mathbf{0.28 ( 0.03 )}$& 0.94 ( 0.10 )&0.73 ( 0.02 )& 1.22 ( 0.07 )& 1.41 ( 0.09 )
\\
20\%
& $\mathbf{0.30 ( 0.03 )}$& 0.85 ( 0.10 )&0.64 ( 0.02 )& 1.16 ( 0.08 )& 1.53 ( 0.13 )\\
30\%
& $\mathbf{0.34 ( 0.03 )}$& 0.76 ( 0.08 )&0.58 ( 0.02 )& 1.07 ( 0.10 )& 1.69 ( 0.11 )\\
40\%
& $\mathbf{0.39 ( 0.03 )}$& 0.76 ( 0.07 )& 0.62 ( 0.03 ) & 1.03 ( 0.10 )& 1.76 ( 0.09 )
\\
50\%
& $\mathbf{0.36 ( 0.03 )}$ & 0.79 ( 0.06 )&0.66 ( 0.03 ) & 1.01 ( 0.10 )& 1.79 ( 0.07 )
\\
70\%
& $\mathbf{0.40 ( 0.03 )}$ & 0.87 ( 0.11 ) & 0.65 ( 0.02 ) & 1.03 ( 0.11 ) & 1.95 ( 0.10 )
\\
90\%
& $\mathbf{1.00 ( 0.03 )}$ & 1.44 ( 0.06 )&1.27 ( 0.06 )& 1.46 ( 0.05 )& 1.91 ( 0.05 )
\\ \bottomrule
\end{tabular}
\caption{\label{t-Chordal U cub}Chordal distance between the true and the estimated score matrix with cubic transformation when $n=200$, $p=20$, $r=4$. Mean (SE) are reported based on 100 replications.}
\end{table*}

\begin{table*}[!t]
\begin{tabular}{@{} c c c c c c@{}}
\toprule
Zero (\%) & \shortstack{truncated\\copula PCA} &  \shortstack{Standard PCA \\(unscaled)} &  \shortstack{Standard PCA \\ (scaled)} & PoissonPCA & ZIPFA  \\ \midrule
0\%
& $\mathbf{0.08 ( 0.02 )}$& 0.65 ( 0.16 )&0.35 ( 0.07 )& 0.65 ( 0.16 )& 1.00 ( 0.01 )
\\
5\%
& $\mathbf{0.08 ( 0.02 )}$& 0.64 ( 0.17 )&0.34 ( 0.07 )& 0.64 ( 0.17 )& 1.00 ( 0.01 )
\\
10\%
& $\mathbf{0.08 ( 0.02 )}$& 0.62 ( 0.17 )&0.33 ( 0.07 )& 0.62 ( 0.17 )& 1.00 ( 0.01 )
\\
15\%
& $\mathbf{0.08 ( 0.02 )}$& 0.61 ( 0.17 )&0.33 ( 0.07 )& 0.61 ( 0.17 )& 1.00 ( 0.01 )
\\
20\%
& $\mathbf{0.08 ( 0.02 )}$& 0.61 ( 0.17 )&0.32 ( 0.07 )& 0.60 ( 0.17 )& 1.01 ( 0.01 )
\\
30\%
& $\mathbf{0.09 ( 0.01 )}$& 0.60 ( 0.16 )&0.32 ( 0.07 )& 0.60 ( 0.16 )& 1.06 ( 0.07 )
\\
40\%
& $\mathbf{0.11 ( 0.02 )}$& 0.60 ( 0.16 )&0.33 ( 0.07 )& 0.60 ( 0.16 )& 1.42 ( 0.12 )
\\
50\%
& $\mathbf{0.13 ( 0.02 )}$& 0.62 ( 0.16 )&0.35 ( 0.07 )& 0.62 ( 0.16 )& 1.47 ( 0.10 )
\\
70\%
& $\mathbf{0.19 ( 0.02 )}$& 0.77 ( 0.15 )&0.49 ( 0.10 )& 0.77 ( 0.15 )& 1.58 ( 0.12 )
\\
90\%& $\mathbf{0.45 ( 0.05 )}$& 1.15 ( 0.09 )&0.96 ( 0.12 )& 1.15 ( 0.09 )& 1.72 ( 0.07 )
\\ \bottomrule
\end{tabular}
\caption{\label{t-Chordal V exp}Chordal distance between the true and the estimated loading matrix with exponential transformation when $n=200$, $p=20$, $r=4$. Mean (SE) are reported based on 100 replications.}
\end{table*}

\begin{table*}[!t]
\begin{tabular}{@{} c c c c c c@{}}
\toprule
Zero (\%) & \shortstack{truncated\\copula PCA} & \shortstack{Standard PCA\\(unscaled)}&  \shortstack{Standard PCA\\(scaled)} & PoissonPCA & ZIPFA  \\ \midrule
0\%
& $\mathbf{0.21 ( 0.04 )}$& 0.90 ( 0.15 )&0.73 ( 0.08 )& 1.04 ( 0.12 )& 1.02 ( 0.03 )
\\
5\%
& $\mathbf{0.21 ( 0.04 )}$& 0.88 ( 0.15 )&0.72 ( 0.08 )& 1.02 ( 0.12 )& 1.02 ( 0.01 )
\\
10\%
& $\mathbf{0.22 ( 0.04 )}$& 0.87 ( 0.15 )&0.70 ( 0.08 )& 1.01 ( 0.12 )& 1.03 ( 0.01 )
\\
15\%
& $\mathbf{0.22 ( 0.03 )}$& 0.85 ( 0.15 )&0.69 ( 0.08 )& 0.99 ( 0.12 )& 1.04 ( 0.01 )
\\
20\%
& $\mathbf{0.23 ( 0.03 )}$& 0.84 ( 0.15 )&0.68 ( 0.08 )& 0.98 ( 0.12 )& 1.05 ( 0.01 )
\\
30\%
& $\mathbf{0.24 ( 0.03 )}$& 0.82 ( 0.15 )&0.66 ( 0.08 )& 0.95 ( 0.13 )& 1.11 ( 0.08 )
\\
40\%
& $\mathbf{0.26 ( 0.03 )}$& 0.82 ( 0.15 )&0.66 ( 0.07 )& 0.94 ( 0.13 )& 1.54 ( 0.16 )
\\
50\%
& $\mathbf{0.29 ( 0.03 )}$& 0.83 ( 0.15 )& 0.68 ( 0.07 ) & 0.95 ( 0.12 )& 1.68 ( 0.12 )
\\
70\%
& $\mathbf{0.40 ( 0.03 )}$ & 1.00 ( 0.13 ) & 0.85 ( 0.08 ) & 1.08 ( 0.11 )& 1.95 ( 0.05 )
\\
90\%
& $\mathbf{1.00 ( 0.03 )}$ & 1.47 ( 0.05 ) & 1.40 ( 0.06 ) & 1.48 ( 0.05 ) & 1.89 ( 0.06 )
\\ \bottomrule
\end{tabular}
\caption{\label{t-Chordal U exp}Chordal distance between the true and the estimated score matrix with exponential transformation when $n=200$, $p=20$, $r=4$. Mean (SE) are reported based on 100 replications.}
\end{table*}

Besides the scaling and mixed transformation, we compare the performance of our truncated copula PCA with the competing PCA methods using cubic and exponential transformations. The data generation mechanism follows Section~3.2 if the main paper. Tables \ref{t-Chordal V cub}-\ref{t-Chordal U cub} show average chordal distance for loadings and scores, respectively, across 100 replications for the cubic transformation as a function of zero inflation proportion. Tables \ref{t-Chordal V exp}-\ref{t-Chordal U exp} show average chordal distance for loadings and scores, respectively, across 100 replications for the exponential transformation as a function of the zero inflation proportion. The results are similar to those in the main paper. In terms of both loading matrix and score estimation, the proposed method always has the smallest errors across all zero-proportion settings. 

\newpage
\subsection{Additional results for application on microbiome data}\label{app:extraSJ}
As shown in the main paper, the second principal component from the proposed truncated copula PCA and the first principal component from ZIPFA are statistically significantly associated with neutropenia at $10\%$ significance level.     
Tables \ref{t-logistic tcPCA Neutropenia 2PC}-\ref{t-logistic ZIPFA Neutropenia 2PC} show the details of these two logistic regression models.  Table \ref{t-logistic tcPCA Fever 2PC} shows the logistic regression model for association between fever and the first two principal components of the truncated copula PCA.

\begin{table}[!t]
\centering
\begin{tabular}{@{} c c c c c  @{}}
\toprule
 & Estimate & Std. Error & odds ratio & p-values \\ \midrule
PC1 & -0.02& 0.086& 0.980& 0.814
\\
PC2 & 0.203& 0.110& 1.226& $\mathbf{0.065}$
\\
race & -0.666& 0.600& 0.514& 0.267
\\
age & -0.086& 0.054& 0.917& 0.110
\\
gender & -0.641& 0.473& 0.527& 0.175
\\
risk & 0.286& 0.512& 1.331& 0.577
\\
\bottomrule
\end{tabular}
\caption{\label{t-logistic tcPCA Neutropenia 2PC} Logistic regression model for association between neutropenia in the chemotherapy induction courses and the first two principal component scores derived from truncated copula PCA using gut microbiome families before chemotherapy, adjusting for age, race, gender and risk group.}
\end{table}

\begin{table}[!t]
\centering
\begin{tabular}{@{} c c c c c  @{}}
\toprule
 & Estimate & Std. Error & OR & p-values \\ \midrule
PC1 & 0.104 & 0.058 & 1.109 & $\mathbf{0.072}$ \\
PC2 & 0.044 & 0.030 & 1.045 & 0.142 \\
race & -0.588 & 0.584 & 0.556 & 0.314 \\
age & -0.064 & 0.053 & 0.938 & 0.228 \\
gender & -0.683 & 0.477 & 0.505 & 0.153 \\
risk & 0.311 & 0.517 & 1.365 & 0.547 \\
\bottomrule
\end{tabular}
\caption{\label{t-logistic ZIPFA Neutropenia 2PC} Logistic regression model for association between neutropenia in the chemotherapy induction courses and the first two principal component scores derived from ZIPFA using gut microbiome families before chemotherapy, adjusting for age, race, gender and risk group.}
\end{table}

No models found significant associations with bacteremia or diarrhea, the predictor p-values for corresponding models are shown in Tables \ref{t-pvalues for logistic models for bacteremia w 2PCs}-\ref{t-pvalues for logistic models for Diarrhea w 2PCs}

\begin{table}[!t]
\centering
\begin{tabular}{@{} c c c c c  @{}}
\toprule
 & Estimate & Std. Error & odds ratio & p-values \\ \midrule
PC1 & -0.039 & 0.086 & 0.962 & 0.649 \\
PC2 & 0.254 & 0.112 & 1.289 & $\mathbf{0.024}$ \\
race & -0.62 & 0.587 & 0.538 & 0.291 \\
age & -0.104 & 0.054 & 0.901 & $\mathbf{0.054}$ \\
gender & -0.522 & 0.472 & 0.594 & 0.269 \\
risk & 0.142 & 0.508 & 1.153 & 0.779 \\
\bottomrule
\end{tabular}
\caption{\label{t-logistic tcPCA Fever 2PC} Logistic regression model  for association between fever in the chemotherapy induction courses and the first two principal component scores derived from truncated copula PCA using gut microbiome families before chemotherapy, adjusting for age, race, gender and risk group.}
\end{table}

\begin{table}[!t]
\centering
\begin{tabular}{@{} c c c c c c @{}}
\toprule
 & \shortstack{truncated\\ copula PCA}  & \shortstack{standard PCA\\(unscaled)} & \shortstack{standard PCA\\(scaled)} & Poisson PCA & ZIPFA \\ \midrule
PC1 & 0.986& 0.972& 0.949& 0.945& 0.920\\
PC2 & 0.957& 0.440& 0.674& 0.395& 0.699
\\
race & 0.677& 0.821& 0.747& 0.843& 0.696
\\
age & 0.961& 0.979& 0.985& 0.969& 0.992
\\
gender & 0.504& 0.579& 0.555& 0.595& 0.562
\\
risk & 0.540& 0.563& 0.560& 0.564& 0.600\\
\bottomrule
\end{tabular}
\caption{\label{t-pvalues for logistic models for bacteremia w 2PCs} Logistic regression model p-values for association between bacteremia in the chemotherapy induction courses and the first two principal component scores from truncated copula PCA and the comparing methods, adjusting for age, race, gender and risk group before chemotherapy. Associations are considered statistically significant at the 10\% significance level ($p < 0.1$) and results are bolded. }
\end{table}

\begin{table}[!t]
\centering
\begin{tabular}{@{} c c c c c c @{}}
\toprule
 & \shortstack{truncated\\ copula PCA}  & \shortstack{standard PCA\\ (unscaled)} & \shortstack{standard PCA\\ (scaled)} & \shortstack{Poisson PCA} & \shortstack{ZIPFA} \\ 
 \midrule
PC1 & 0.943& 0.628& 0.657& 0.663& 0.907
\\
PC2 & 0.971& 0.610& 0.468& 0.569& 0.850\\
race & 0.409& 0.524& 0.537& 0.528& 0.437
\\
age & 0.215& 0.244& 0.235& 0.241& 0.213
\\
gender & 0.375& 0.424& 0.430& 0.424& 0.426
\\
risk & 0.886& 0.893& 0.901& 0.892& 0.944
\\
\bottomrule
\end{tabular}
\caption{\label{t-pvalues for logistic models for Diarrhea w 2PCs} Logistic regression model p-values for association between diarrhea in the chemotherapy induction courses and the first two principal component scores from truncated copula PCA and the comparing methods, adjusting for age, race, gender and risk group before chemotherapy. Associations are considered statistically significant at the 10\% significance level ($p < 0.1$) and results are bolded. }
\end{table}

\clearpage

\newpage

\begin{figure}[!t]
\centering
    \begin{minipage}[b]{.75\textwidth}
    \includegraphics[width=1\textwidth]{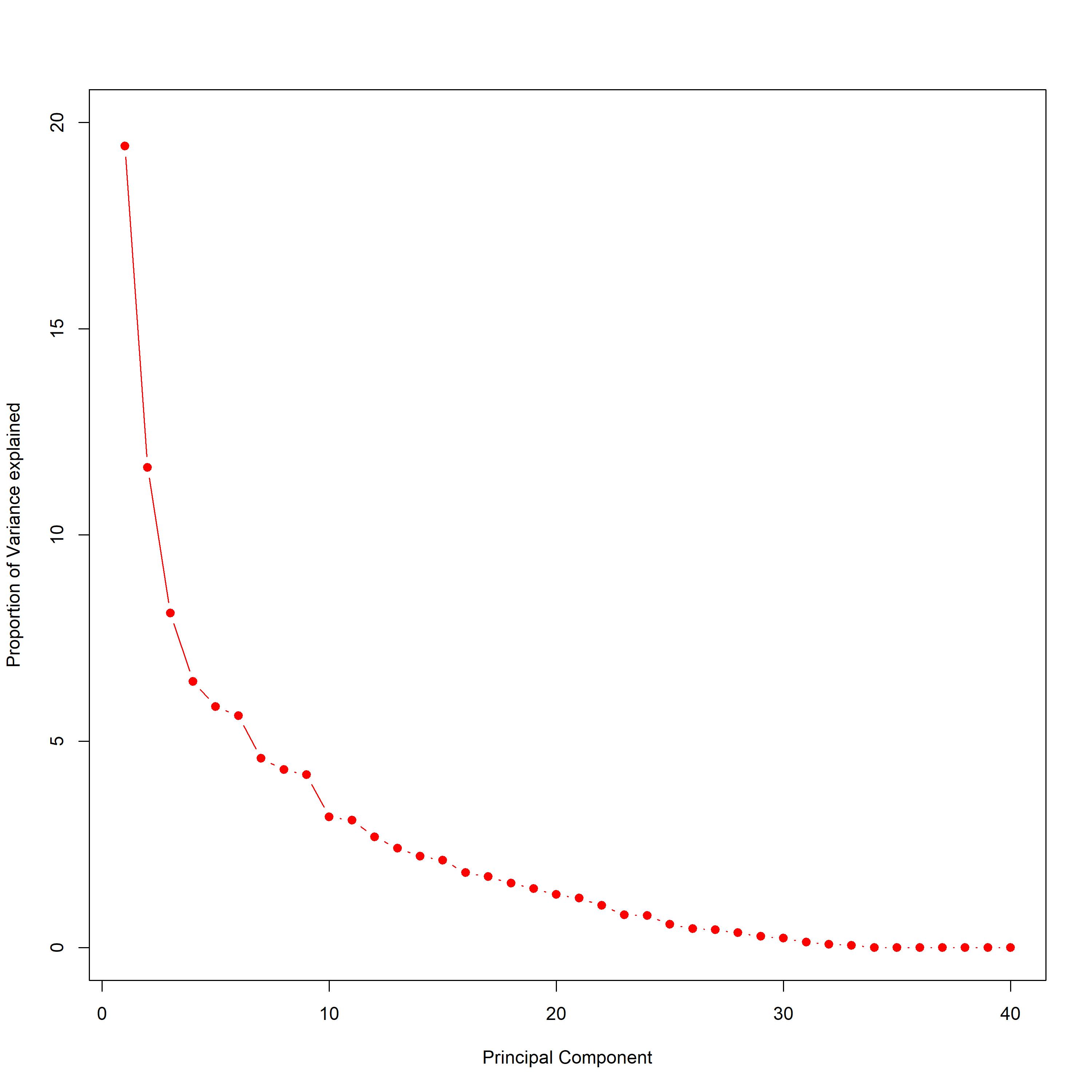}
	\caption{Scree plot with truncated copula PCA.}
    \label{fig:scree_tcPCA}
    \end{minipage}
    \hfill
\end{figure}

Figure~\ref{fig:scree_tcPCA} shows the scree plot for the estimated latent correlation matrix $\widehat \Sigma$ using truncated copula PCA. Based on the position of the "elbow", we have also considered models for fever and neutropenia based on five rather than two principal components for each method. Table~\ref{tab:AIC} shows AIC values from each model, demonstrating that model based on two PCs from the proposed truncated copula PCA method fits best.

\begin{table}[!t]
\centering
\begin{tabular}{p{2.5cm}p{2.5cm}p{2.5cm}p{2.5cm}p{2.5cm}}
\toprule
 \textbf{Model} & \textbf{Neutropenia (2PC)}& \textbf{Neutropenia (5PC)}& \textbf{Fever$\quad$ (2PC)}& \textbf{Fever$\quad$ (5PC)} \\ \midrule
truncated copula PCA  & \textbf{131.93}& 136.97& \textbf{132.38}& 138.17\\
standard PCA (unscaled) & 134.98& 138.29& 137.01& 137.76\\
standard PCA (scaled) & 134.28& 133.43 & 135.31& 134.33\\
Poisson PCA & 134.98& 138.16& 136.97& 137.37\\
ZIPFA & 132.25& 133.52& 135.71& 136.77\\

\bottomrule
\end{tabular}
\caption{\label{tab:AIC} AIC values for five models using 2 vs 5 principal components (PCs) for predicting Neutropenia and Fever. Lowest AICs within each outcome are in bold.}
\end{table}

\end{document}